\documentclass[runningheads]{llncs}

\usepackage{lipsum}
\usepackage{listings}
\usepackage{subfig}
\usepackage{caption}
\usepackage{subcaption}
\usepackage{booktabs}

\usepackage[T1]{fontenc}
\usepackage{graphicx}

\begin{document}
\title{YNote: A Novel Music Notation for Fine-Tuning LLMs in Music Generation}
\titlerunning{YNote: A Novel Music Notation}
\author{Shao-Chien Lu\inst{1} \and
Chen-Chen Yeh\inst{1} \and
Hui-Lin Cho\inst{1} \and
Chun-Chieh Hsu\inst{1} \and
Tsai-Ling Hsu{2}\and
Cheng-Han Wu\inst{3} \and
Timothy K. Shih\inst{3}\and
Yu-Cheng Lin\inst{1}
}
\authorrunning{S.C. Lu et al.}

\institute{Department of Computer Science and Engineering,\\
Yuan Ze University, Taoyuan, Taiwan
\email{\{s1111534,s1111511,s1112003,s1101545\}@mail.yzu.edu.tw} 
\email{linyu@saturn.yzu.edu.tw} \and
International Bachelor Program in Informatics,\\
Yuan Ze University, Taoyuan, Taiwan\\
\email{s1093530@mail.yzu.edu.tw}\and
Department of Computer Science and Information Engineering,\\
National Central University, Taoyuan, Taiwan\\
\email{\{whoshank,timothykshih\}@gmail.com}}
\maketitle              
\begin{abstract}
The field of music generation using Large Language Models (LLMs) is evolving rapidly, yet existing music notation systems, such as MIDI, ABC Notation, and MusicXML, remain too complex for effective fine-tuning of LLMs. These formats are difficult for both machines and humans to interpret due to their variability and intricate structure. To address these challenges, we introduce \textbf{YNote}, a simplified music notation system that uses only four characters to represent a note and its pitch. YNote’s fixed format ensures consistency, making it easy to read and more suitable for fine-tuning LLMs.

In our experiments, we fine-tuned GPT-2 (124M) on a YNote-encoded dataset and achieved BLEU and ROUGE scores of 0.883 and 0.766, respectively. With just two notes as prompts, the model was able to generate coherent and stylistically relevant music. We believe YNote offers a practical alternative to existing music notations for machine learning applications and has the potential to significantly enhance the quality of music generation using LLMs.

\end{abstract}


\section{Introduction}
Recently, Large Language Models (LLMs) have made significant advances. By fine-tuning pre-trained LLMs with task-specific datasets, we can generate desired outputs across various domains. For instance, previous work \cite{geerlings2020interacting} has used ABC notation \cite{ABCNotation} to interact with GPT-2 \cite{radford2019language} for music generation, yielding impressive results. However, ABC notation is difficult for humans to read due to its complex structure and lack of a fixed format, which makes fine-tuning LLMs using ABC notation challenging.

Other data formats for music representation, such as MIDI, ABC Notation, and MusicXML, are also used, but they are not sufficiently simple. To address these challenges, we introduce a new data structure called \textbf{YNote}. YNote offers a simplified music notation that is both human-readable and structured in a fixed format, making it ideal for fine-tuning LLMs. The details of the YNote format are described in Section \ref{IntroductionOfYNote}. Section \ref{Methodology} introduces our methodology, and experimental results are evaluated using BLEU and ROUGE scores in Section \ref{ExperimentsAndResults}.


\section{Related Work}
In this section, we introduce three common music notation systems—MIDI, MusicXML, and ABC Notation—and use \textit{Boat on Tai Lake}, a Chinese folk tune, as an example to demonstrate each format.

\begin{figure}[htbp]
    \centering
    \subfloat[Sheet Music]{
        \begin{minipage}[t]{\textwidth}
            \centering
            \includegraphics[width=\textwidth]{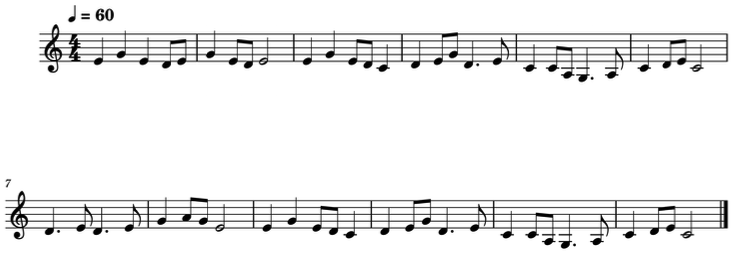}  
        \end{minipage}
    } \quad
    \centering
    \subfloat[MIDI (ASCII Code Representation)]{
        \begin{minipage}[b]{0.46\textwidth}
\begin{lstlisting}[basicstyle=\fontsize{7}{9}\selectfont\ttfamily]
MFile 1 1 480
MTrk
0 Tempo 500000
0 KeySig 0 major
0 TimeSig 4/4 24 8
0 Meta TrkName "Taihu Boat"

0 On ch=1 n=62 v=100
480 Off ch=1 n=62 v=100
480 On ch=1 n=67 v=100
960 Off ch=1 n=67 v=100
960 On ch=1 n=71 v=100
1440 Off ch=1 n=71 v=100
1440 On ch=1 n=69 v=100
...
12000 Off ch=1 n=65 v=100
12000 On ch=1 n=64 v=100
12480 Off ch=1 n=64 v=100
12480 Meta TrkEnd
\end{lstlisting}
        \end{minipage}
    } \quad
    \subfloat[MusicXML]{
        \begin{minipage}[b]{0.47\textwidth}
\begin{lstlisting}[basicstyle=\fontsize{6}{8}\selectfont\ttfamily, language=XML]
<?xml version="1.0" encoding="UTF-8"?>
<museScore version="3.02">
  <programVersion>3.6.2</programVersion>
  <Score>
    ...
    <TimeSig>
      <sigN>4</sigN>
      <sigD>4</sigD>
    </TimeSig>
    <Tempo>
      <tempo>1.667</tempo>
      <text>
        <sym>metNoteQuarterUp</sym>=100
      </text>
    </Tempo>
    <Chord>
      <durationType>quarter</durationType>
      <Note>
        <pitch>64</pitch>
        <tpc>18</tpc>
        <velocity>90</velocity>
        <veloType>user</veloType>
      </Note>
    </Chord>
    ...
  </Score>
</museScore>
\end{lstlisting}
        \end{minipage}
    }
    \qquad
    \subfloat[ABC Notation]{
        \begin{minipage}[b]{0.47\textwidth}
\begin{lstlisting}[basicstyle=\fontsize{7}{9}\selectfont\ttfamily]
X: 1
T: Boat on Tai Lake
C: Chinese Jiangnan Sizhu Folk Music
M: 4/4
L: 1/8
Q: 1/4=100
K: C % 0 sharps
V: 1
E2 G2 E2 DE| G2 ED E4| E2 G2 ED C2|
D2 EG2<D2E| C2 CA,2<G,2A,| C2 DE C4|
D3E2<D2E| G2 AG E4| E2 G2 ED C2| 
D2 EG2<D2E| C2 CA,2<G,2A,| C2 DE C4|
\end{lstlisting}
        \end{minipage}
    } \quad
    \subfloat[YNote]{
        \begin{minipage}[b]{0.46\textwidth}
\begin{lstlisting}[basicstyle=\fontsize{7}{9}\selectfont\ttfamily]
E504G504E504D508E508
G504E508D508E502
E504G504E508D508C504
D504E508G508D54.E508
C504C508A408G44.A408
C504D508E508C502
D54.E508D54.E508
G504A508G508E502
E504G504E508D508C504
D504E508G508D54.E508
C504C508A408G44.A408
C504D508E508C502
\end{lstlisting}
        \end{minipage}
    }
    \caption{\textit{Boat on Tai Lake} in Various Music Notations}
    \label{fig:code-table}
\end{figure}

\subsection{MIDI}
MIDI (Musical Instrument Digital Interface) \cite{de2017understanding} is a standardized protocol for communication between musical devices, first developed in 1983 by several electronic music equipment manufacturers. Unlike audio signals, MIDI transmits digital control information, enabling electronic instruments, computers, synthesizers, and audio equipment to communicate with one another. Its primary purpose is to send instructions, such as which notes to play, their velocity, and timbre, guiding devices on how to generate sound.

The MIDI protocol specifies a transmission rate of 31.25 kbps, which is fast enough to ensure low-latency signal transmission necessary for real-time musical control. MIDI data is generally divided into three main categories: Channel Messages, System Messages, and Control Change Messages.

\begin{itemize}
    \item Channel Messages convey essential performance information, such as note-on and note-off events, note velocity, and other parameters needed for playing back notes.
    \item System Messages handle global controls, including timing synchronization and system resets.
    \item Control Change Messages are used to make real-time adjustments to sound characteristics, such as pitch bends, modulation, and pedal controls.
\end{itemize}

MIDI technology is widely used across various fields, including music production, live performance control, sheet music generation, music education, film scoring, and sound design. It is also integral to automated and interactive music systems. Thanks to its efficient data transmission and standardized protocol, MIDI allows musicians, producers, and sound designers to easily compose music, control effects, and synchronize digital audio elements, providing powerful support for both music creation and performance. Figure 1\_b shows an example of a MIDI file for \textit{Boat on Tai Lake} represented in ASCII code.


\subsection{MusicXML}
MusicXML (Music Extensible Markup Language) \cite{MAKEMUSIC,MUSICNOTATIONCOMMUNITYGROUP} is a standard open format based on XML, creating a universal format for common Western musical notation used for the interchange and distribution of digital scores. Musical information can be utilized by score programs, sequencers, other performance applications, music education software, and music databases. The hierarchical structure of MusicXML reflects the structure of music itself — from higher-level elements like pieces and sections, down to lower-level details like notes and beats.

MusicXML has two top-level elements, representing the partwise and timewise score formats, which are determined by the root element. The top-level element also defines the structural framework of the lower-level elements. If the root element is <score-partwise>, the musical part is the primary structure, containing multiple <part> elements, each of which includes several <measure> elements. On the other hand, if the root element is <score-timewise>, the measure is the primary structure, containing multiple <measure> elements, each of which includes several <part> elements.

The actual music in the score is represented by lower-level elements, each of which contains a group of music data. These elements are composed of zero or more <note>, <backup>, <direction>, <attributes>, <sound>, and other elements. Sub-elements like <key> and <time> within <attributes>, as well as <pitch> within <note>, define the basic structure of each measure and the specific details of the notes.

The score header refers to the elements located at the top of the score, providing basic information and settings about the piece. This includes the title, composer, arranger, performer, copyright details, and more. It is typically composed of elements like <work>, <movement-number>, <movement-title>, <part-list>, and others.

Overall, MusicXML provides a standardized format that is widely supported by many applications and offers high readability for developers. Its ability to thoroughly describe elements like notes, harmony, and rhythm allows for a more complete representation of the score. However, the detailed information also leads to larger file sizes, and in some cases, the structure may be overly complex or verbose. Additionally, MusicXML can be more difficult for musicians to understand. Figure 1\_c shows an example of MusicXML for \textit{Boat on Tai Lake}.


\subsection{ABC Notation}
ABC Notation \cite{ABCNotation} is a simple, text-based music notation system that is relatively easy for humans to read. Over the past three decades, it has been used to represent tens of thousands of musical pieces, which are commonly referred to as ABC tunes.

An ABC tune consists of two main parts: the tune header and the tune body. The tune header contains basic information about the music, including reference number, title, composer, origin, region, meter, unit note length, tempo, parts, transcription, key, rhythm, and other metadata such as background information and instructions.

The tune body records the actual musical content. Notes are represented by their English names ("A", "B", "C", etc.), and octaves are indicated with commas for lower octaves and apostrophes (or lowercase letters) for higher octaves. For example, a "C" in a lower octave is written as "C,", while a "C" in a higher octave is written as "C'" or "c".

The note duration is defined by the L value in the tune header. For instance, if L is set to 1/8, then "C" represents an eighth note, "C/2" represents a sixteenth note, and "C2" represents a quarter note.

Bars are separated by the "|" symbol, and repeated sections can be indicated using ":|". Additionally, "K" and "V" symbols are used to represent key signatures and clefs, and custom symbols can be defined by the user.

Although ABC Notation is more readable than traditional sheet music, its flexible and varied representations can lead to the same piece being notated in multiple ways. This flexibility, along with the presence of optional information, makes it difficult to parse automatically with high accuracy. An example of ABC Notation is shown in Figure 1\_d.


\section{Introduction of YNote} 
\label{IntroductionOfYNote}
YNote is a new data structure designed to represent sheet music, consisting of
two key components: the pitch and duration of notes.

\subsection{Pitch}
In YNote, pitch is represented by an English note name followed by a single digit. The English note names are: C, D, E, F, G, A, and B—also known as Do, Re, Mi, Fa, Sol, La, and Si. Notes with a half-step are represented by the corresponding lowercase letter. For example, "C$\sharp$" or "D$\flat$" is denoted as "c". The digit indicates the octave, where the number 4 represents middle C (C4), while a lower and higher octave are denoted as C3 and C5, respectively.

We define the frequency of A4 as 440 Hz, and the frequency of each subsequent note is calculated based on the equal temperament system. To maintain a simple and consistent format, we use "00" to represent a rest.

\subsection{Duration}
Duration in YNote is represented with two characters. For example, "01" denotes a whole note, "02" represents a half note, "04" is a quarter note, and so on. Dotted, double-dotted, and triplet notes are also represented using two characters. A dotted note adds a period after the note length (e.g., a dotted quarter note is written as "4."). Double-dotted notes add a colon (e.g., "4:"), and triplets are indicated by adding the digit 3 (e.g., "43"). When encountering dotted sixteenth, thirty-second, or sixty-fourth notes, we avoid using "16." since it would require three characters and violate our two-character limit. Instead, we use "S," "T," and "U" to represent these notes, respectively. For example, a dotted sixteenth note is written as "S." and a double-dotted thirty-second note as "T:". This system ensures that all notes are represented using exactly two
characters.

We use the tick as a unit, defining a quarter note in a 4/4 time signature as 480 ticks and a whole note as 1920 ticks. This system allows for an easy conversion of the YNote format into a waveform audio file, enabling playback of the music.

\begin{figure}[!h]
\centering
\includegraphics[width=\textwidth]{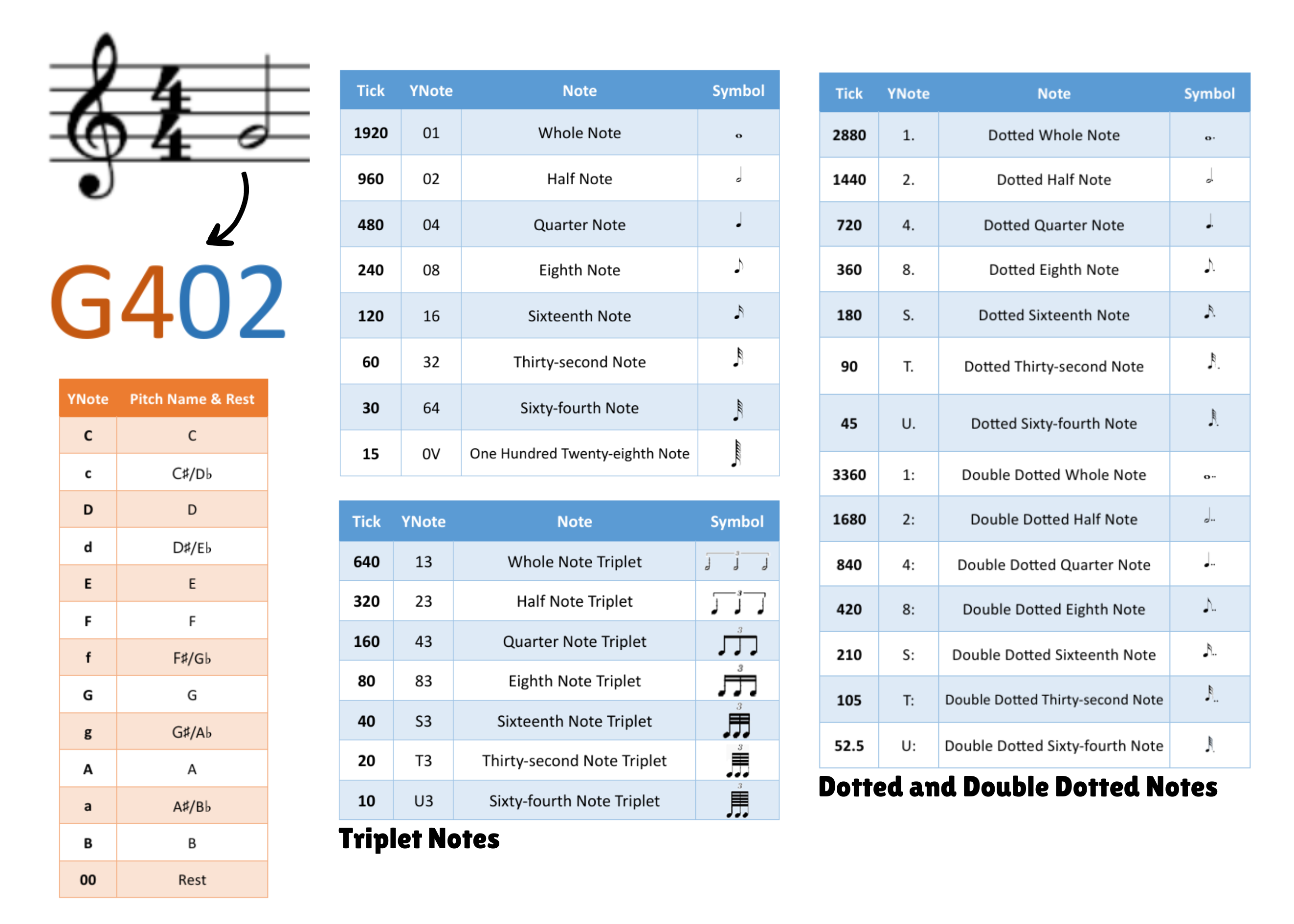}
\caption{Overview of YNote Format}
\label{fig:ynote}
\end{figure}

With this system, most musical notations can be captured in YNote format. Although some special notes cannot be represented, their occurrence is rare enough to be excluded. YNote’s fixed format is both human-readable due to its simplicity and well-suited for machine learning because of its structured consistency. Figure 1\_e shows an example of YNote notation used to represent \textit{Boat on Tai Lake}, and Figure \ref{fig:ynote} provides an overview of the YNote format. We can easily combine pitch and duration to represent a note and convert a YNote into standard notation.

\section{Methodology}
\label{Methodology}
In this section, we describe how YNote-encoded music was used to fine-tune the GPT-2 model and generate music in a similar style.

\subsection{Background}
\textbf{Music Generation} is the process of creating music using algorithms and machine learning models, and it has become a popular research topic in recent years. Various methods can be used for music generation. For example, \cite{geerlings2020interacting} fine-tuned the GPT-2 model using music in ABC Notation, achieving impressive results, while \cite{banar2022systematic} fine-tuned GPT-2 using MIDI data and carefully designed metrics. Even diffusion models, such as those described in \cite{huang2023noise2music}, have been used to generate music.

\textbf{GPT-2} (Generative Pre-trained Transformer 2) \cite{radford2019language} is a large-scale, unsupervised language model developed by OpenAI, trained on 8 million web pages. It is built on the Transformer architecture \cite{vaswani2017attention}, which leverages self-attention mechanisms to model dependencies between input and output tokens. GPT-2 is pre-trained on a large corpus of text data and can be fine-tuned for specific tasks to generate high-quality text outputs. There are several versions of GPT-2, ranging from 124M to 1.5B parameters, with the 124M model being the smallest, which we used in our experiments.

\subsection{Fine-tuning of GPT-2}
\begin{figure}[]
    \subfloat[Fine-Tuning Step]{%
      \includegraphics[width=0.45\textwidth]{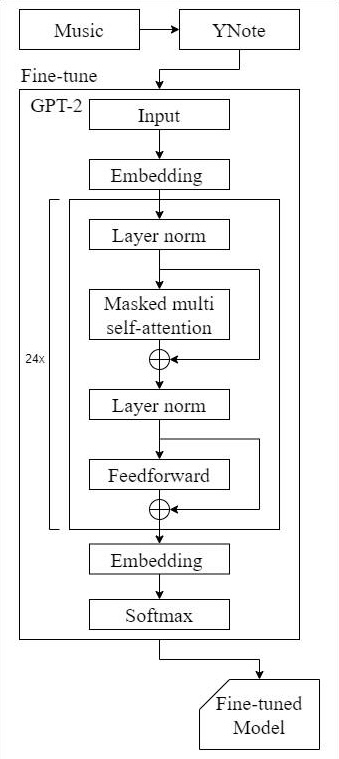}}
    \subfloat[Inference Step]{%
      \includegraphics[width=0.45\textwidth]{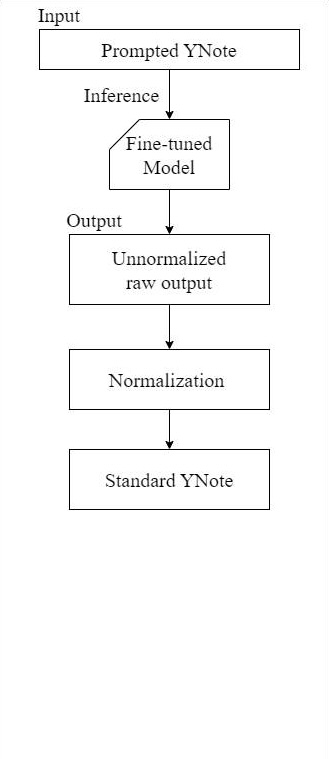}
    }
    \caption{Flowchart for Generating Music in YNote Format}
    \label{fig:3_pipeline}
\end{figure}

Figure \ref{fig:3_pipeline} illustrates the step of fine-tuning GPT-2 and generating music in a similar style. The architecture of the pre-trained GPT-2 model is based on \cite{heilbron2019tracking}. In the fine-tuning step, we first convert the music into YNote format and then fine-tune the pretrained GPT-2 model using PyTorch. During the inference step, we provide the fine-tuned GPT-2 model with prompts, such as the first bar or the first and last notes of each bar, to generate new music. However, sometimes the YNote format generated by GPT-2 is invalid. Therefore, we must normalize it to conform to the standard YNote format.

\clearpage


\section{Experiments and Results} 
\label{ExperimentsAndResults}
In this section, we describe how YNote-encoded music was used to fine-tune the GPT-2 model \cite{radford2019language}, and present the corresponding BLEU \cite{papineni2002bleu} and ROUGE \cite{lin2004rouge} scores.

\subsection{Dataset Description}
We converted 190 pieces of Jiangnan-style music into the YNote format. The prompt consisted of the first and last notes of each bar to guide GPT-2 in learning and generating music. Additionally, we experimented with using the entire first bar as the prompt, which also yielded satisfactory results.

\subsection{Dataset Description}
Due to hardware constraints and speed requirements, we selected GPT-2 (124M) as the base model. The data was split, reserving 30\% as the test set, and five pieces of music were withheld from both the training and test sets for evaluating BLEU and ROUGE scores. The AdamW optimizer \cite{loshchilov2017decoupled} was used with a learning rate of 5e-5, a warmup of 100 steps, and training over 20 epochs.

\subsection{Results}
We fine-tuned GPT-2 on a single NVIDIA GeForce RTX 4070 Ti Super, and the training took approximately half an hour. To evaluate the model, we generated five pieces of music for each of the reserved pieces and calculated their corresponding BLEU and ROUGE scores.

BLEU measures how well the generated text overlaps with a reference text, using n-grams. Its values range from 0 to 1, with 1 indicating perfect overlap and 0 indicating no overlap. ROUGE is also used to assess the quality of machine-generated text, with values similarly ranging from 0 to 1, where higher is better. Both metrics are commonly used for evaluating the performance of language models.

In our results (see Figure \ref{tbl:results_1} and \ref{tbl:results_2}), we observe that the music generated using our method closely resembles the original compositions, even when the prompt is not part of the training set. This outcome aligns with our expectations. However, on some occasions, the GPT-2-generated music in YNote format does not fully adhere to the intended structure, requiring preprocessing to correct these errors. Specifically, we modified 125 out of 7569 characters (approximately 1.6\%) in the first bar and 149 out of 6485 characters (about 2.2\%) when focusing on the first and last notes of the prompt. Additionally, Figure \ref{fig:6_music_sheet} and \ref{fig:7_music_sheet} shows some examples of music generated by GPT-2.

\begin{figure}%
  \centering
  \subfloat[][BLEU Scores]{
    \begin{tabular}{|l||l|l|l|l|}
    \hline
             & 1-gram & 2-gram & 3-gram & 4-gram \\ \hline
    Sample 1 & 0.736  & 0.375  & 0.099  & 0.033  \\ \hline
    Sample 2 & 0.818  & 0.438  & 0.153  & 0.046  \\ \hline
    Sample 3 & 0.799  & 0.229  & 0.054  & 0.013  \\ \hline
    Sample 4 & 0.469  & 0.253  & 0.202  & 0.168  \\ \hline
    Sample 5 & 0.667  & 0.493  & 0.314  & 0.219  \\ \hline
    \end{tabular}
  }%
  \qquad
  \subfloat[][ROUGE Scores]{
    \begin{tabular}{|l||l|l|}
    \hline
             & 1-gram & 2-gram \\ \hline
    Sample 1 & 0.529  & 0.226  \\ \hline
    Sample 2 & 0.652  & 0.294  \\ \hline
    Sample 3 & 0.544  & 0.177  \\ \hline
    Sample 4 & 0.400  & 0.250  \\ \hline
    Sample 5 & 0.655  & 0.449  \\ \hline
    \end{tabular}
  }
  \caption{Qualitative Evaluation of Fine-Tuned GPT-2 Models Using the First Bar as Prompt}%
  \label{tbl:results_1}%
\end{figure}

\begin{figure}%
  \centering
  \subfloat[][BLEU Scores]{
    \begin{tabular}{|l||l|l|l|l|}
    \hline
             & 1-gram & 2-gram & 3-gram & 4-gram \\ \hline
    Sample 1 & 0.713  & 0.405  & 0.155  & 0.035  \\ \hline
    Sample 2 & 0.838  & 0.590  & 0.200  & 0.033  \\ \hline
    Sample 3 & 0.852  & 0.472  & 0.150  & 0.027  \\ \hline
    Sample 4 & 0.750  & 0.407  & 0.153  & 0.040  \\ \hline
    Sample 5 & 0.883  & 0.576  & 0.297  & 0.132  \\ \hline
    \end{tabular}
  }%
  \qquad
  \subfloat[][ROUGE Scores]{
    \begin{tabular}{|l||l|l|}
    \hline
             & 1-gram & 2-gram \\ \hline
    Sample 1 & 0.649  & 0.313  \\ \hline
    Sample 2 & 0.714  & 0.258  \\ \hline
    Sample 3 & 0.723  & 0.275  \\ \hline
    Sample 4 & 0.666  & 0.319  \\ \hline
    Sample 5 & 0.766  & 0.407  \\ \hline
    \end{tabular}
  }
  \caption{Qualitative Evaluation of Fine-Tuned GPT-2 Models with the First and Last Notes of Each Bar as Prompt}%
  \label{tbl:results_2}%
\end{figure}

\clearpage

\begin{figure}[]
    \subfloat[Sample 1]{%
      \includegraphics[width=0.9\textwidth]{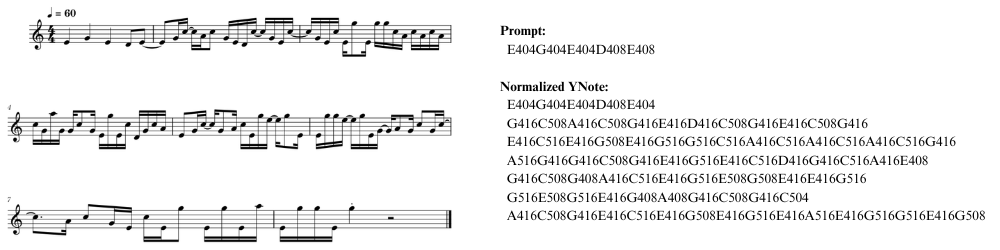}
    } \\
    \subfloat[Sample 2]{%
      \includegraphics[width=0.9\textwidth]{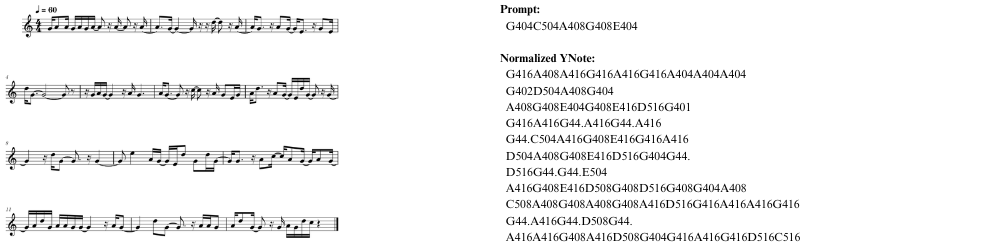}
    } \\
    \subfloat[Sample 3]{%
      \includegraphics[width=0.9\textwidth]{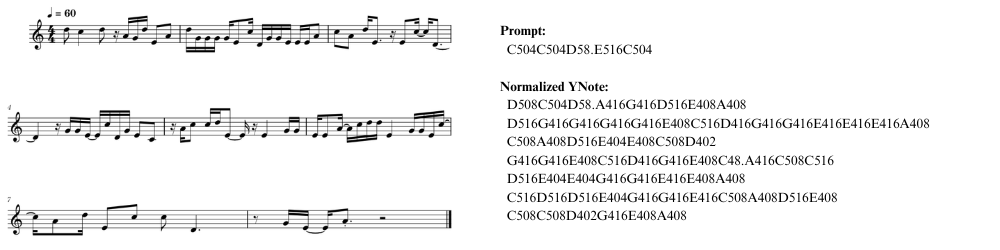}
    } \\
    \subfloat[Sample 4]{%
      \includegraphics[width=0.9\textwidth]{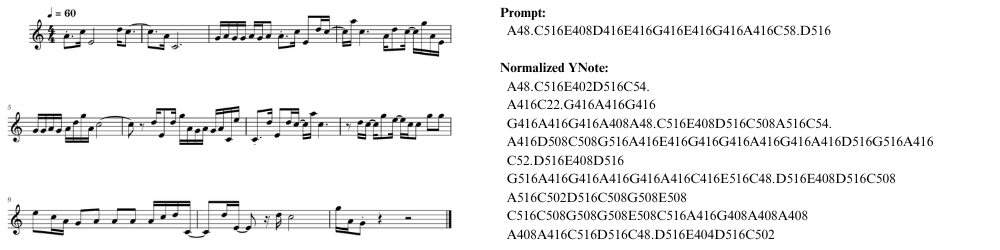}
    } \\
    \subfloat[Sample 5]{%
      \includegraphics[width=0.9\textwidth]{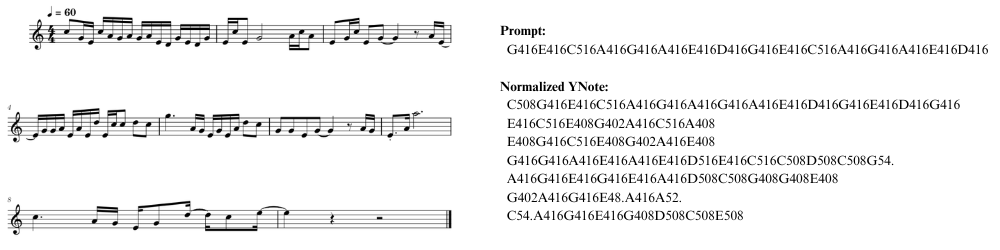}
    }
    \caption{The Music Generated by Fine-Tuned GPT-2 Using the First Bar as Prompt}
    \label{fig:6_music_sheet}
\end{figure}

\begin{figure}[]
    \subfloat[Sample 1]{%
      \includegraphics[width=0.9\textwidth]{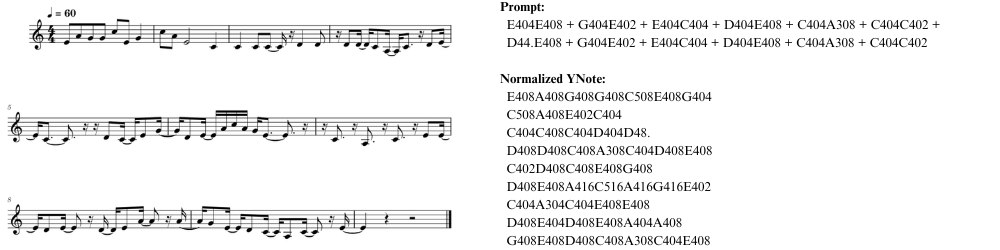}
    } \\
    \subfloat[Sample 2]{%
      \includegraphics[width=0.9\textwidth]{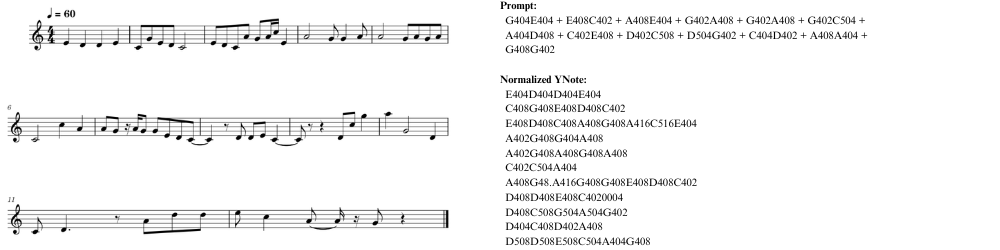}
    } \\
    \subfloat[Sample 3]{%
      \includegraphics[width=0.9\textwidth]{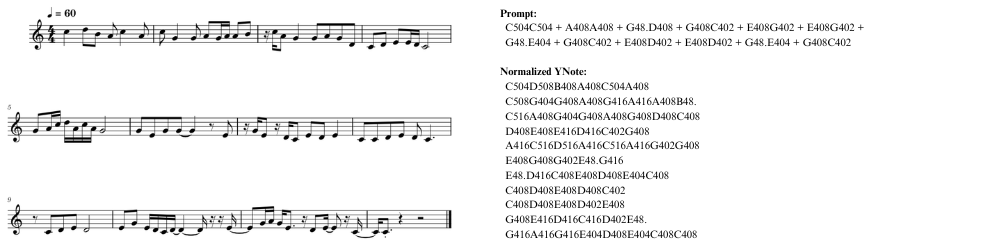}
    } \\
    \subfloat[Sample 4]{%
      \includegraphics[width=0.9\textwidth]{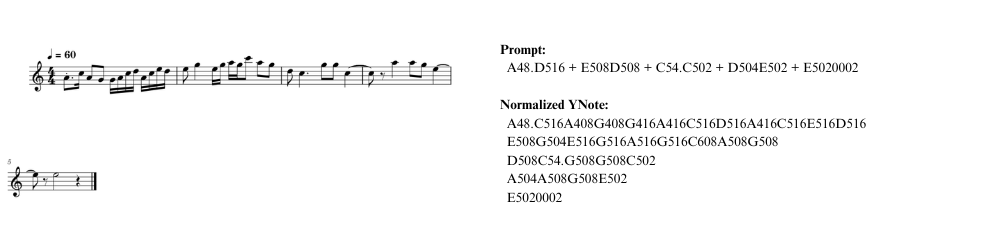}
    } \\
    \subfloat[Sample 5]{%
      \includegraphics[width=0.9\textwidth]{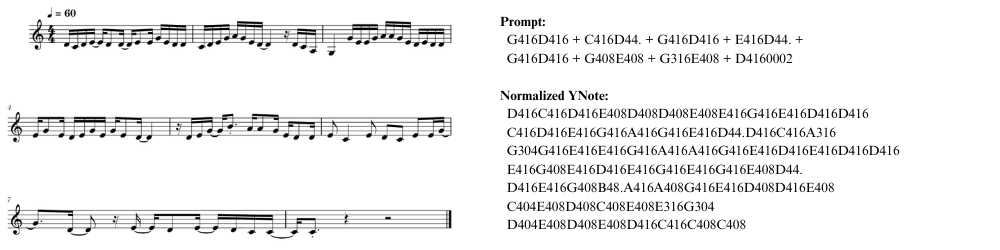}
    }
    \caption{The Music Generated by Fine-tuned GPT-2 with the First and Last Notes of Each Bar as Prompt}
    \label{fig:7_music_sheet}
\end{figure}

\clearpage


\section{Conclusion}
In this paper, we introduced a new data structure for representing sheet music. To simplify the format, we excluded certain uncommon notes and situations. This makes the structure more readable and suitable for machine learning, thanks to its fixed format. Our approach achieved BLEU and ROUGE scores of 0.883 and 0.766, respectively. Additionally, we demonstrated the ability to generate music in specific styles based on given notes. We believe that the YNote format offers a promising alternative for recording music information, with potential for further development.

Additionally, GPT-2’s robust network allows us to easily generate music in different styles by preparing the corresponding dataset in YNote format and feeding it into the model. This is made possible due to the structured and easily understandable nature of the YNote format.


\bibliographystyle{splncs04}
\bibliography{YNote_A_Novel_Music_Notation}

\end{document}